\documentclass[useAMS,fleqn,usenatbib]{mnras}
\usepackage{newtxtext,newtxmath}
\usepackage{multirow}
\usepackage[T1]{fontenc}
\usepackage{ae,aecompl}
\usepackage{todonotes}
\usepackage{graphicx}	
\usepackage{subfigure}
\usepackage{amsmath}	
\usepackage{threeparttable}
\usepackage{booktabs}
\usepackage[normalem]{ulem}

\def\apj {ApJ}
\def\apjl {ApJL}
\def\apjs {ApJS}
\def\aj {AJ}
\def\aap {A\&A}
\def\mnras {MNRAS}

\def\nat {Nature}

\def\R200 {R_{200}}

\def\Gyr {\rm Gyr}

\title[ROGER]
{ROGER: Reconstructing Orbits of Galaxies in Extreme Regions using machine 
learning techniques}

\author[de los Rios et al.]{\parbox[t]{\textwidth}{
Mart\'in de los Rios,$^1$\thanks{E-mail: martindelosrios13@gmail.com}
H\'ector J. Mart\'inez,$^{2,3}$
Valeria Coenda,$^{2,3}$ Hern\'an Muriel,$^{2,3}$
Andr\'es N. Ruiz,$^{2,3}$ Cristian A. Vega-Mart\'inez,$^{4,5}$ and Sof\'ia A. Cora$^{6,7}$}
\\
\\
$^{1}$ICTP South American Institute for Fundamental Research \& Instituto de F\'isica Te\'orica, Universidade Estadual \\ Paulista, 01140-070, S\~ao Paulo-SP, Brazil
\\ 
$^{2}$Instituto de Astronom\'ia Te\'orica y Experimental (CCT C\'ordoba, CONICET, UNC), 
Laprida 854, X5000BGR, C\'ordoba, Argentina\\
$^{3}$Observatorio Astron\'omico, Universidad Nacional de C\'ordoba,
Laprida 854, X5000BGR, C\'ordoba, Argentina\\
$^{4}$Instituto de Investigaci\'on Multidisciplinar en Ciencia y Tecnolog\'ia, Universidad de La Serena, Ra\'ul Bitr\'an 1305, La Serena, Chile\\
$^{5}$Departamento de Astronom\'ia, Universidad de La Serena, Av. Juan Cisternas 1200 Norte, La Serena, Chile\\
$^{6}$Instituto de Astrof\'isica de La Plata (CCT La Plata, CONICET, UNLP), 
Observatorio Astron\'omico, Paseo del Bosque S/N, B1900FWA,\\
La Plata, Argentina\\
$^{7}$Facultad de Ciencias Astron\'omicas y Geof\'isicas, Universidad Nacional de La Plata, 
Observatorio Astron\'omico, \\
Paseo del Bosque S/N, B1900FWA, La Plata, Argentina
}

\date{Accepted XXX. Received YYY; in original form ZZZ}

\pubyear{2020}

\begin{document}
\label{firstpage}
\pagerange{\pageref{firstpage}--\pageref{lastpage}}
\maketitle

\begin{abstract}
We present the \texttt{ROGER} (Reconstructing Orbits of Galaxies in Extreme Regions) 
code, which uses three different  machine learning techniques to 
classify galaxies in, and around, clusters, according to their projected 
phase-space position. We use a sample of 34 massive,  
$M_{200}>10^{15} h^{-1} M_{\odot}$, galaxy clusters in the 
MultiDark Planck 2 (MDLP2) simulation at redshift zero. 
We select all galaxies with stellar mass $M_{\star} \ge 10^{8.5} 
h^{-1}M_{\odot}$, as computed by the 
semi-analytic model of galaxy formation SAG, that are 
located in, and in the vicinity of, these clusters and classify 
them according
to their orbits. We train \texttt{ROGER} to retrieve the original classification
of the galaxies from their projected phase-space 
positions.
For each galaxy, \texttt{ROGER} gives as output the probability of being a cluster
galaxy, a galaxy that has recently fallen into a cluster, 
a backsplash galaxy, an infalling galaxy, or an interloper. 
We discuss the performance 
of the machine learning methods and potential uses of our code.
Among the different methods explored, we find the K-Nearest Neighbours 
algorithm achieves the best performance. 
\end{abstract}

\begin{keywords}
galaxies: clusters: general -- galaxy: haloes -- galaxies: kinematics and dynamics -- methods: numerical -- methods: analytical
\end{keywords}

\section{Introduction}
\label{sect:intro}
Galaxies in the Universe show a wide variety of properties as a result of the 
action of both, internal and environmental processes. Clusters of galaxies 
constitute the most extreme environments in the Universe for galaxy evolution. 
They are the most massive objects $(\sim10^{14-15} M_{\odot})$ in virial 
equilibrium, are characterised by a deep gravitational potential well, a large
number of galaxy members, and an intracluster medium filled with hot
ionised gas. Galaxies in clusters exhibit different properties compared to 
galaxies that reside in the field, or in less massive systems. 

Several physical processes affect galaxies inside clusters in a simultaneous 
way. One of these mechanisms is the ram pressure stripping 
(e.g. \citealt{GG:1972,Abadi:1999,Book:2010,Steinhauser:2016}). This process can
remove an important fraction of the cold gas from galaxies, resulting in the 
inhibition of star formation. Although this mechanism is more effective at the
central regions of massive clusters, it has been reported in less massive systems
(e.g. \citealt{Rasmussen:2006,Jaffe:2012,Hess:2013}). 
Ram pressure stripping occurs as galaxies move at high speeds through the hot 
ionised gas of the intracluster medium, which collides with the cold gas of the
galaxies and removes it. The warm gas from the galactic halo can also be removed 
by the gas of the intracluster medium, a process known as starvation (e.g. 
\citealt{Larson:1980,Balogh:2000,McCarthy:2008,Bekki:2009,Bahe:2013,
Vijayaraghavan:2015}). 
This process can cut off further gas cooling from the galaxy's halo gas that fuels future star formation.
\citet{Kawata:2008} predicted that starvation can act in galaxy groups as well.
Another physical process that works on galaxies in their passage through the 
deep potential well of the cluster is tidal stripping (e.g. 
\citealt{Zwicky:1951,Gnedin:2003a,Villalobos:2014}).
It can induce a central star formation burst \citep{Byrd:2001}, bar 
instabilities \citep{Lokas:2016}, changes in the pattern of the spiral arms 
\citep{Semczuk:2017}, and truncate dark matter haloes (e.g. 
\citealt{Gao:2004,Limousin:2009}). 
In the outskirts of clusters, mechanisms like galaxy-galaxy interaction, known 
as harassment, are more effective (e.g. 
\citealt{Moore:1996,Moore:1998,Gnedin:2003b,Smith:2015}). 
Most of the processes mentioned above tend to 
decrease or to completely suppress the star 
formation in galaxies. As a consequence, galaxies in clusters are typically 
red, early-type, with an old stellar population, 
and have little or none star formation at all. 

Clusters of galaxies are continually accreting galaxies. Some galaxies may 
fall as members of galaxy groups and, therefore, they may 
have already experienced environmental effects that 
accelerated the consumption of their gas reservoir prior to 
entering the cluster. 
This is known as pre-processing (e.g. 
\citealt{Mihos:2004,Fujita:2004}), and has both observational 
and theoretical support (e.g. \citealt{Balogh:1999,McGee:2009,deLucia:2012,
Jaffe:2012,Wetzel:2013,Hou:2014}). In the outskirts of clusters, 
not only star forming and pre-processed galaxies are found, 
but also backsplash galaxies. 
These are galaxies that have orbited the central regions of the cluster only 
once since their infall. They are currently outside the cluster, and will fall 
back to the cluster in the future (e.g. 
\citealt{Gill:2005,Rines:2005,Aguerri:2010,Muriel:2014}). Backsplash galaxies
do not necessarily have become passive yet. 
They have suffered the extreme 
environmental effects of the inner regions of a 
cluster, which  has surely 
left traces in their physical properties.
Consequently, the characterisation of this population of 
galaxies is important to understand the effects that the cluster environment produces in 
galaxies.

Since the efficiency of the physical 
processes described above depends 
heavily on the history of galaxies in and around clusters, 
a detailed knowledge of galaxy orbits is essential. 
It is customary to classify galaxies 
around clusters using different criteria according to their position in the 
\emph{Projected Phase-Space Diagram} (PPSD). This two-dimensional
space combines the projected cluster-centric distance, with the 
line-of-sight velocity relative to the cluster.
\citet{Oman:2013} use a N-body simulation to compile a catalogue of the orbits 
of satellite haloes in cluster environments. They found that satellite haloes 
in different phases of their orbits occupy different regions in the PPSD.
\citet{Mahajan:2011} quantify the 
decrease of star formation in backsplash galaxies. \citet{Muzzin:2014} find that
quiescent, star forming and post-starburst galaxies in clusters at $z\sim 1$, 
are distributed differently in the PPSD. \citet{Muriel:2014} study the properties
of galaxies in the outskirts of a sample of 90 galaxy clusters. They split 
galaxies into two classes: those with high relative velocity, and those with 
low relative velocity, being the latter candidates to be backsplash galaxies.
These authors find that backsplash candidates are systematically older,
redder, and have formed fewer stars in the last 3 Gyrs than 
high-velocity galaxies. In addition, \citet{HF:2014} and 
\citet{Jaffe:2015} 
infer the orbital histories of cluster galaxies from their PPSD positions to investigate the effects of 
ram pressure on the gas fraction of galaxies. \citet{OmanHudon:2016} measure
quenching time-scales in clusters as a function of the position in the PPSD.
\citet{Yoon:2017} trace the gas stripping histories of galaxies infalling into
the Virgo cluster using a reference sample in the PPSD. \citet{Jaffe:2018} 
reconstruct the stripping history of jellyfish galaxies using their PPSD position
as an indication of their orbits. 

Using cosmological hydrodynamic N-body simulations of groups and clusters,
\citet{Rhee:2017} separate galaxies in the PPSD as a function of the time 
elapsed since their infall into the system: first, recent, intermediate, and 
ancient infallers. They define regions in the PPSD where each of these types of
galaxies are more likely to be found. These regions 
can, in turn, be used to classify galaxies 
from their PPSD position. An alternative tool is given by 
\citet{Pasquali:2019}. They use cosmological simulations of groups and clusters 
and derive zones of constant mean infall time. They use these zones to study 
the environmental effects upon satellite galaxies. 
They provide an analytical form
for the curves that define each zone. \citet{Smith:2019} use these zones in the 
PPSD to create samples of ancient and recent infallers 
among satellite galaxies in a SDSS group catalogue. 
They use these samples to study the stellar 
mass growth histories of galaxies as a function of infall time. 

Although many interesting results have been obtained  
using the PPSD, in 
practice it is very difficult to determine with certainty
whether a particular galaxy is a backsplash, or it is infalling to the cluster 
for the first time, or if it has already become a cluster member. 
In this work we take a different approach by classifying galaxies relating their 
three-dimensional orbits with their position in the PPSD. We use a sample of
massive clusters from the MultiDark Planck 2 (MDLP2) cosmological simulation 
\citep{klypin_mdpl2_2016} and generate the galaxy population with the 
semi-analytic model of galaxy formation SAG \citep{cora_sag_2018}.
We  classify galaxies in and around these clusters into five types, according 
to their three-dimensional orbits. Then, we develop a Machine Learning 
code and train it to recover the orbital classification (3-D) of the galaxies 
out of their PPSD position (2-D). Machine learning techniques represent a new 
way of analysing big data-sets in an agnostic and homogeneous way. Taking into
account the amount of data generated by current and future surveys and 
simulations, the data-driven techniques will become a fundamental tool 
for their analysis. 
These methods are very useful and powerful tools to find 
patterns and relations between the variables that are involved in a specific 
problem. In particular, these methods are especially good in classification
problems \citep{messi, clecio, cora}. 

This  article  is  organised  as  follows:  we  describe the data sets of 
simulated clusters and galaxies in Sect. \ref{sect:data}, where we also define 
different galaxy types according to their orbits; in sect. \ref{sect:roger}, 
we present our code
\texttt{ROGER} (Reconstructing Orbits of Galaxies in Extreme Regions)
that relates the 
two-dimensional PPSD position of galaxies to their 3-D orbits, and analyse its
performance; finally, we present our conclusions in Sect. \ref{sect:conclu}.

\begin{figure}
    \centering
    \includegraphics[width=1.1\columnwidth]{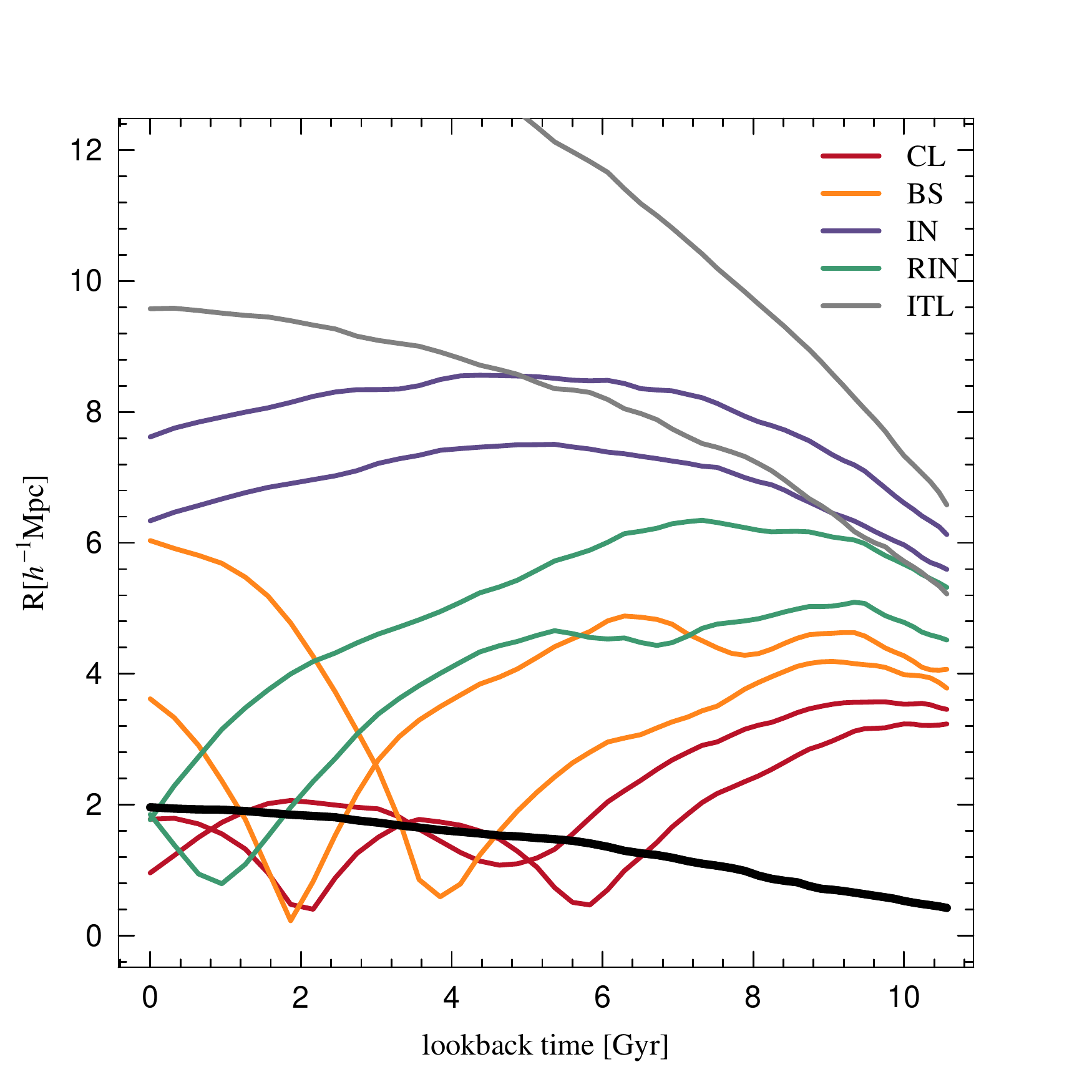}
    \caption{Cluster-centric distance as a function of lookback time for ten 
    galaxies in/around one of the clusters in our sample. The thick black 
    solid line represents the $R_{200}$ of the cluster. Different colours 
    correspond to galaxies classified as: cluster members 
    (CL, red lines), recent infallers (RIN, green lines), backsplash (BS, orange 
    lines), infall galaxies (IN, violet lines), and interlopers (ITL, gray 
    lines).}
    \label{fig:orbit}
\end{figure}

\section{The sample of simulated clusters and galaxies}
\label{sect:data}
We use a sample of clusters and galaxies from a simulated galaxy 
catalogue constructed using the semi-analytic model of galaxy formation and 
evolution SAG ({\em Semi-Analytic Galaxies}, \citealt{cora_sag_2018}). As a 
backbone for this synthetic catalogue, we use dark matter haloes and subhaloes, 
and their corresponding merger trees, extracted from the cosmological simulation 
MDPL2 \citep{klypin_mdpl2_2016}. In this section, we briefly present the MDPL2 
simulation and  the
SAG code, and describe the samples of simulated clusters and 
galaxies we use throughout the paper.

\subsection{The MDPL2 simulation}
The MDPL2 simulation is part of the \textsc{MultiDark} suite of dark matter 
simulations, publicly available at the \textsc{CosmoSim} 
database\footnote{https://www.cosmosim.org/} 
\citep{riebe_multidark_2013,klypin_mdpl2_2016}. The simulation has $3840^3$ dark
matter particles in a comoving cubic volume of $1~h^{-3} {\rm Gpc^3}$, evolved 
from redshift 120 to redshift 0. The cosmology adopted corresponds to a 
$\Lambda$CDM model with parameters consistent with  Planck measurements 
\citep{planck_cosmology_2014,planck_cosmology_2016}: 
$\Omega_{\rm m} = 0.307$, $\Omega_{\rm \Lambda}=0.693$, $\Omega_{\rm b}=0.048$, 
$\sigma_8 = 0.823$, $n=0.96$, and $h=0.678$.
Dark matter haloes were identified using the \textsc{Rockstar} halo finder 
\citep{behroozi_rockstar_2013}, keeping all the dark matter bounded structures 
with at least 20 particles. The final catalogue comprises $\sim 127 \times 10^6$ 
haloes, whose merger trees were constructed using the \textsc{ConsistentTrees} 
algorithm \citep{behroozi_trees_2013}.

\subsection{The SAG model}
The version of SAG used in this work was presented in detail in 
\citet{cora_sag_2018}. This model includes the main physical processes relevant to 
galaxy formation and evolution: radiative cooling of hot gas in central and 
satellite galaxies, star formation in quiescent and bursty modes, being the 
latter triggered by galaxy mergers and disc instabilities, detailed treatment of 
chemical enrichment of gaseous and stellar components, supernova feedback and 
stellar winds, gas ejection and reincorporation of hot gas, central 
super-massive 
black hole growth and AGN feedback, ram pressure and tidal stripping. For a
complete and detailed description of all of these processes and their
implementations, we refer the reader to \citet{cora_sag_2006}, 
\citet{lagos_sag_2008}, \citet{tecce_sag_2010}, \citet{gargiulo_sag_2015},
\citet{munnozarancibia2015}, \citet{ruiz_sag_2015}, \citet{cora_sag_2018}, 
\citet{collacchioni_sag_2018}, and \citet{cora_sag_2019}.

\subsection{Galaxies in and around clusters in the MDPL2-SAG catalogue}
From the MDPL2 simulated volume, we select all haloes at redshift zero that 
have a mass, computed within the region that encloses 200 times the critical 
density, $M_{\rm 200}\ge 10^{15}h^{-1}M_{\odot}$. Furthermore, we impose an 
isolation criterion by requiring that they have no companion haloes more massive 
than $0.1\times M_{\rm 200}$, within $5\times R_{\rm 200}$, where
$R_{\rm 200}$ is the radius enclosing the overdensity. With this restriction,
we exclude from our analysis haloes undergoing a major merger, or interacting 
with a massive companion. Both these situations are likely to affect galaxy 
orbits to a great extent in the vicinity of the main halo.
Out of the 85 haloes more massive than $10^{15}h^{-1}M_{\odot}$
in the MDPL2 volume, our selection results in a set of 34 massive, relaxed
haloes that constitute our cluster sample for the present paper. 

In order to guarantee completeness,
we impose a low stellar mass cut-off of $\log_{10}(M_{\star}^{\rm min}/ 
h^{-1}M_{\odot})=8.5$ in our sample of galaxies 
\citep{Knebe2018}. Below this mass limit, observed stellar mass 
functions are not well reproduced and galaxy properties are not followed in a 
reliable way.

For each cluster in our sample, we follow the trajectory of all central 
galaxies and of those satellite galaxies that keep their dark matter 
substructure\footnote{Orphan galaxies are avoided in this analysis.}
and that end up in a region that includes the cluster and its surroundings.
We choose each of these regions to be a cylinder elongated along the $z-$axis
of the simulation box to include not only the cluster galaxies and galaxies in 
the surroundings of the cluster, but also interlopers, i.e., galaxies that will 
appear in or around the cluster in projection but are unrelated to it; 
interlopers constitute the main source of contamination in the PPSD.
There is no lost of generality by choosing cylinders parallel to 
the $z-$axis, i.e., we are choosing the line-of-sight direction to be 
this axis of the simulation box. The dimensions of these cylinders are: a 
radius of $5\times R_{\rm 200}$, and a longitude in the $z-$axis that extends 
as far as to include all galaxies within $|\Delta V_z+H_0\Delta z|\le 3\sigma$,
where $\Delta V_z$ is the galaxy peculiar velocity in the $z$ direction relative
to the cluster, $\Delta z$ is the proper distance between the galaxy and the 
cluster in the $z-$direction, $H_0=67.8\,{\rm km}\,{\rm s}^{-1}\,{\rm Mpc}^{-1}$ 
is the Hubble constant, and $\sigma$ is the one dimensional velocity dispersion
of the cluster. As shown by \citet{Munari:2013}, the measurement of $\sigma$
produces different values depending on whether dark matter particles or 
subhaloes are used in the computation. For consistency, we computed $\sigma$ 
out of the satellite galaxies more massive than $M_{\star}^{\rm min}$,
using the biweight estimator of \citet{Beers:1990}. We recall that satellite 
galaxies in the MDPL2-SAG catalogue are the central galaxies of the subhaloes, 
thus our estimation of $\sigma$ is made out of those subhaloes that harbor a 
central galaxy more massive than our chosen stellar mass threshold.

We classify all galaxies in these cylinders into different types according to 
their orbits around clusters. These galaxies are, in turn, used to train and 
test the machine learning algorithms described in the next section. 
We define five classes of galaxies:

\begin{figure}
    \centering
    \includegraphics[width=\columnwidth]{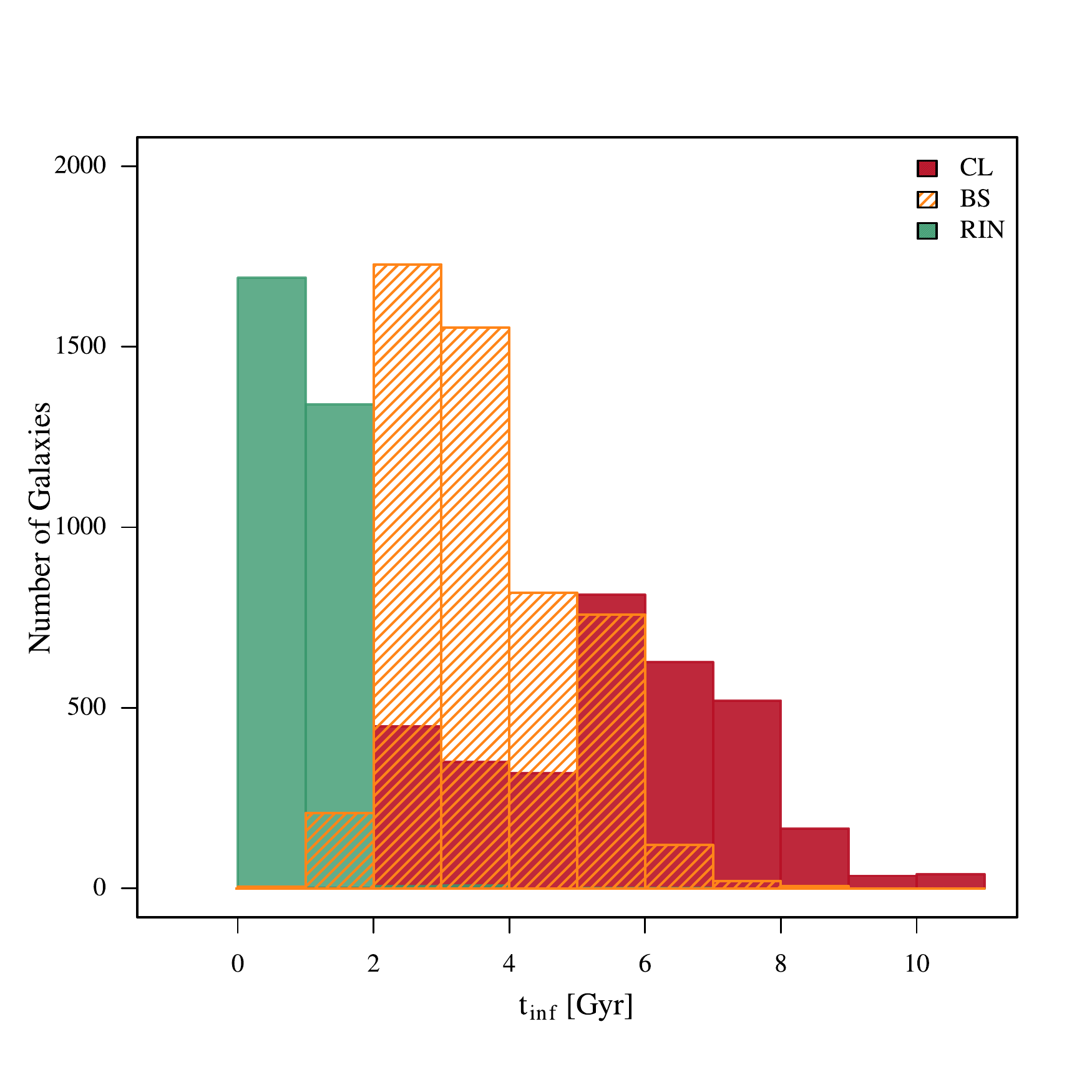}
    \caption{Distribution of the time since first infall for all galaxies in our 
    sample that have been within $R_{\rm 200}$ of a cluster at some point in
    their lifetimes. Colours are as in Fig. \ref{fig:orbit}.}
    \label{fig:times}
\end{figure}

\begin{figure*}
    \centering
    \includegraphics[width=1.\columnwidth]{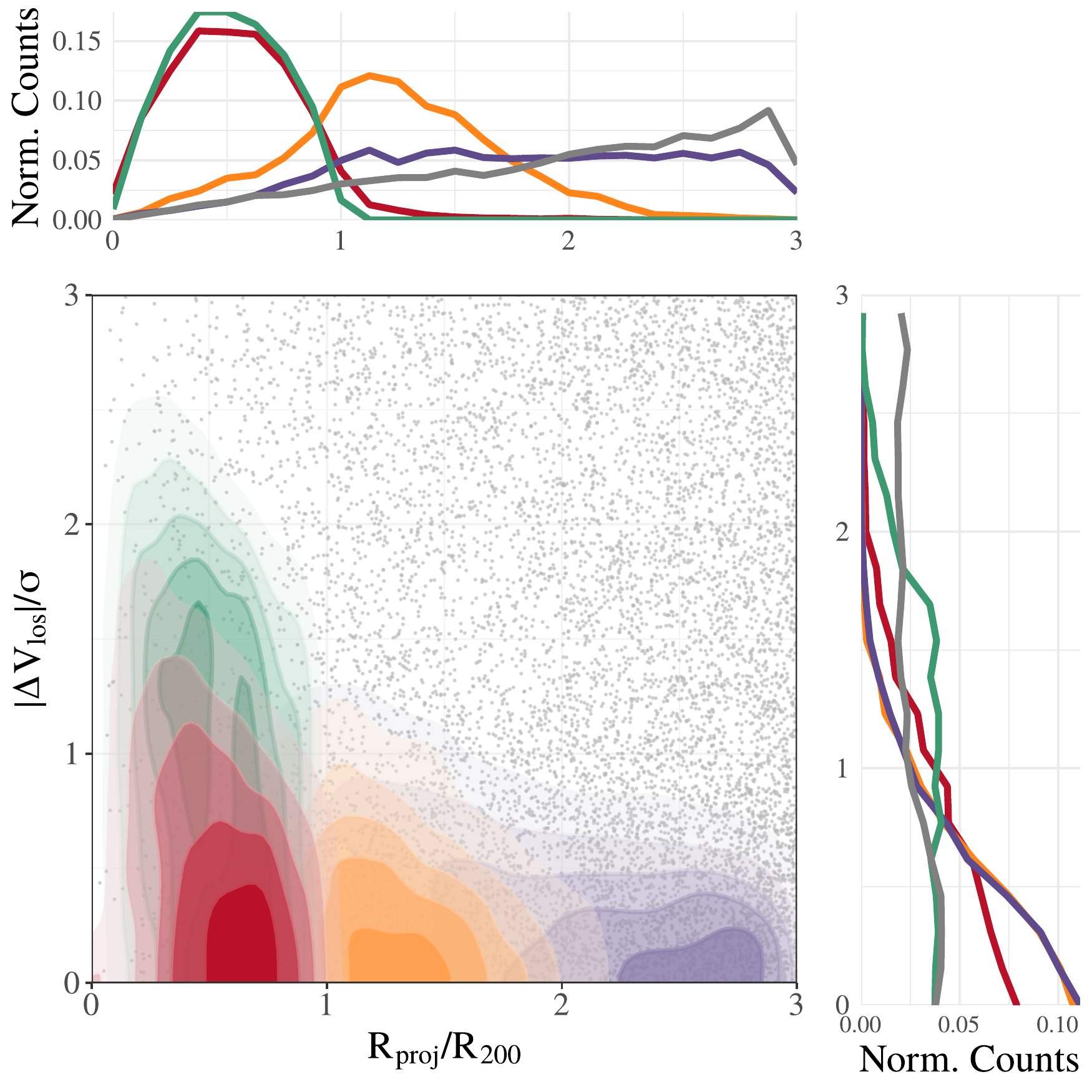}
    \includegraphics[width=1.\columnwidth]{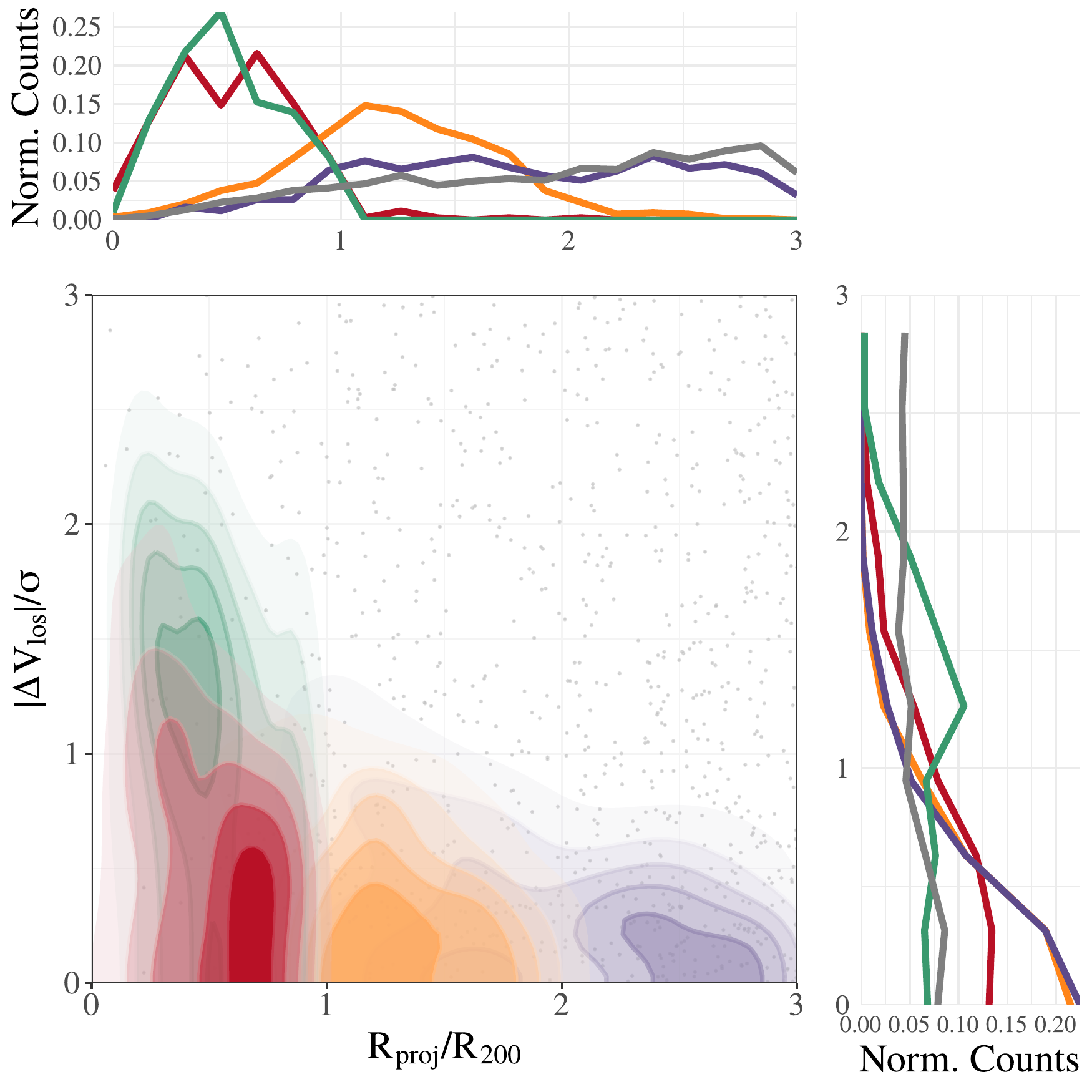}
    \caption{
    Projected phase-space diagram of our full sample (left panel) and of the 
    validation-set (right panel) of MDPL2-SAG galaxies. In the main box, we 
    show in different colour shades the density of galaxies classified as CL
    (red), RIN (green), BS (orange), and IN (violet). ITL galaxies are shown as
    individual gray points for clarity. We show in the upper and 
    right-side panels, the corresponding
    marginalised distributions of $R_{\rm proj}/R_{200}$ 
    and $|\Delta V_{\rm los}|/\sigma$. 
    These distributions are normalised to have unity area.} 
    \label{fig:phase_space}
\end{figure*}

\begin{enumerate}
    \item Clusters members (CL): 
    galaxies that may have crossed $R_{\rm 200}$ several times in the past and 
    now orbit around the cluster centre. Most of them are found within 
    $R_{\rm 200}$ of the cluster centre.
    \item Recent infallers (RIN): galaxies that have crossed $R_{\rm 200}$ only 
    once on their way in in
    the past $2~\Gyr$ (we discuss the choice of this
    timescale below). Among these, there are galaxies that can get further away
    than $R_{\rm 200}$ from the cluster centre in the future. 
    \item Backsplash galaxies (BS): galaxies that have crossed $R_{\rm 200}$ 
    exactly 
    twice. The first time on their way in, and the second time
    on their way out of the cluster, where they are found now. Most of these 
    galaxies will fall back into the cluster. 
    Accordingly to this definition,
    some RIN may become BS in the future.
    \item Infalling galaxies (IN): galaxies that have never been closer than 
    $R_{\rm 200}$ to the cluster centre, and their main halo\footnote{In the 
    simulation, the main halo is the galaxy's own halo for a central galaxy, 
    and it is the central galaxy's halo for a satellite galaxy.} has negative
    radial velocity relative to the cluster. We consider these galaxies to 
    be either bounded to the cluster, or will potentially fall into the cluster 
    in the future.
    \item Interlopers (ITL): galaxies that have never been closer than $R_{200}$ 
    to the cluster centre, but unlike IN, their main halo has a positive radial 
    velocity relative to the cluster, i.e., the halo is receding away from the 
    cluster at redshift zero. In contrast to the IN galaxies described 
    above, we consider 
    these galaxies as objects that will not fall into the cluster. They are 
    galaxies unrelated to the cluster that can be confused with classes 
    (i)-(iv) in the PPSD. 
\end{enumerate}

Galaxies of classes (i) to (iv) are of interest in cluster studies, 
while galaxies of class (v) constitute the main source of contamination at 
the time of classifying galaxies in and around clusters from an observational 
point of view.

We show in Fig. \ref{fig:orbit} examples of how our classification scheme works. 
For one of the cluster in our sample, we have chosen examples of galaxies 
from the 
five classes defined above, and show their three dimensional cluster-centric
distance as a function of the lookback time. 

Special attention deserves our choice of 2 Gyr to define RIN galaxies above. 
It makes no sense to classify all galaxies within $R_{\rm 200}$
as cluster members, since some of the galaxies that have entered the clusters
in recent times may get out of the cluster in the future, and become BS galaxies.
On the other hand, it would be very useful to pick out the galaxies that 
are experiencing the effects of the cluster environment for the first time.
Thus, it is important to define a time-scale to tell apart between `old' cluster 
galaxies, and galaxies that have fallen into the cluster not a long time ago.  
When we analysed the distribution of the infall time of galaxies that
are within $R_{\rm 200}$ at redshift zero, we find a clear peak at lookback times
$t_{\rm inf}\le 2$ Gyr; this fact motivates our choice
of the condition imposed to  classify RIN galaxies.
We show in Fig. \ref{fig:times} the distribution of the first infall times of 
galaxies that have been within  $R_{\rm 200}$ at least once in their lifetimes, 
i.e., CL, RIN and BS. 

For real galaxies, only their projected distance to the centre of the cluster,
and their line-of-sight velocity relative to the cluster can be measured, i.e.,
their positions in the PPSD. In Fig. \ref{fig:phase_space} (left panel), we 
show the PPSD of all galaxies in our sample. 
Phase-space positions of 
galaxies relative to their parent cluster's centre are computed by projecting 
the 3D cartesian coordinates of the galaxies in the MDPL2 box into the $(x,y)$ 
plane. Hereafter, the projected distance on this plane, $R_{\rm proj}$, will be 
referred to as the 2D distance, and it will be quoted in units of $R_{200}$ 
unless otherwise specified. On the other hand, the $z-$axis velocity relative to 
the cluster, $\Delta V_{\rm los}\equiv |\Delta V_z+H_0\Delta z|$, will be called 
the line-of-sight velocity, and will 
be quoted in units of the velocity dispersion of the cluster, $\sigma$. 

As can be seen in Fig. \ref{fig:phase_space}, there is much overlap between the 
five classes of galaxies in the PPSD. Thus, deciding how to classify a galaxy 
according to its phase-space position is not trivial. 
CL and RIN galaxies
have similar radial distributions, occupying the region defined by 
$R_{\rm proj}/R_{200}\lesssim 1$. These two classes differ, however, 
in their line-of-sight velocity distributions: CL galaxies are concentrated 
towards $\Delta V_{\rm los}/\sigma \sim 0$, while RIN galaxies show a roughly
flat distribution up to $\Delta V_{\rm los}/\sigma \sim 1.25$, an indication 
that the latter do not constitute a population in virial equilibrium. BS 
galaxies are found preferentially between $0.5R_{200}$ and $2 R_{200}$, 
with a broad peak 
at $\sim 1.2 R_{200}$. Their velocity distribution is very similar to that of 
IN galaxies. These two latter classes are characterised by having typically 
low line-of-sight
velocities. IN galaxies, in turn, have little overlap with CL and RIN galaxies, 
they are mostly located at $R_{\rm proj}/R_{200}>1$, showing a flat radial 
distribution up to $R_{\rm proj}/R_{200}\sim 3$. Finally, ITL galaxies are found 
everywhere in the PPSD, with two distinctive features consistent with a 
population uncorrelated to 
the clusters: their radial distribution is roughly 
linear with $R_{\rm proj}$, and they have an almost flat velocity distribution.
ITL galaxies are a clear source of contamination for BS and IN galaxies, and to 
a much lesser extent for CL and RIN galaxies.

To tackle the problem of classifying galaxies out of their PPSD position, we
explore machine learning techniques in the next section. This constitutes a
new, alternative way to address the problem, not previously found in the 
literature.

\section{Machine Learning classification of galaxies in the 2-D phase space.}
\label{sect:roger}
Machine learning (ML) techniques have proved to be powerful tools for 
classification tasks, as they look for correlations between the input variables, 
also called features, and the classes in which we want to group our data. It is 
important to remark that, in order to achieve this goal, it is mandatory to have 
a reliable dataset in which we must know the input variables (in our case, the 
2D distance to the cluster centre and the relative velocity along the 
line-of-sight), as well as the output classes (in our case, the real orbital 
classification).

Using the dataset described in Sec. \ref{sect:data}, we analyse the performance 
of three different techniques:  K-Nearest Neighbours, Support Vector Machine, 
and Random Forest. We briefly describe them.

\begin{itemize}
    \item K-Nearest Neighbours (KNN): 
    This algorithm estimates the probability of a new object to belong to a 
    certain class taking into account the proportion of neighbours of each 
    class. The neighbours are defined as the $k$ nearest objects
    belonging to the training set. It is worth remarking that as 
    the variables 
    in the $y$ and $x-$axis are normalised and span a similar range of values, 
    we decided to use the Euclidean distance in the PPSD 
    to look for neighbours.
    
    \item Support Vector Machine (SVM, \citealt{svm}): 
    It is a supervised learning algorithm that, given a training set $\{(x_{1}, 
    y_{1}), \ldots , (x_{n}, y_{n})\}$ in a feature space of dimension $d$, it 
    looks for hyper-planes, i.e. hyper-surfaces of dimension $ d-1 $, that
    separate the classes. It is important to note that this algorithm only 
    looks for linear hyper-planes in the feature space. A way to generalise 
    this method to more complex hyper-planes is to enlarge the feature space by 
    means of a set of basis functions $h(x)$. Then the hyper-planes will be 
    searched for in a new feature space defined by $ h(x_{i}) = (h_{1} (x_{i}), 
    h_{2} (x_{i} ), \ldots) $, that translates onto non-linear surfaces in 
    the original feature space.
    
    \item Random Forest (RF): A decision tree algorithm is a supervised
    learning algorithm that subdivides the input feature space into 
    sub-regions and then adjusts a local model to each of these sub-regions. One 
    of the main problems of the decision trees is their instability; this means
    that small changes in the training set can lead to very different 
    predictions. That is why in general it is advisable to train many decision 
    trees and then average the results. 
    RF is an implementation of this technique developed by \citet{rf}. This
    method consists in training $N$ decision trees randomly selecting the
    features that will be studied to sub-divide the input space as explained 
    above. In this way, we can reduce the 
    correlation between different trees. 
\end{itemize}

With the aim of providing a fully consistent and automatic method
for the classification of galaxies, we have created the 
R-package 
\texttt{ROGER} (Reconstructing Orbits of Galaxies in Extreme Regions) that is publicly available through the github repository: 
\href{https://github.com/Martindelosrios/ROGER/}
{https://github.com/Martindelosrios/ROGER/}. 
With this software any user can analyse their own galaxy sample with the methods
described above.

For the analysis of the three machine learning methods, we use the R-package 
\texttt{caret} \citep{caret}. To train the three algorithms described above, 
we first randomly split the full data-set of galaxies into two independent 
samples: a training-set of 26370 galaxies (90 per cent of the total) and a 
validation-set of 2930 galaxies (10 per cent of the total).
Each of these sets is a random pick of the galaxies shown in left panel of
Fig. \ref{fig:phase_space}. For a better comparison, we show in the right panel 
of Fig. \ref{fig:phase_space} the PPSD of the galaxies in the validation-set.
It can be seen that this sub-sample follows similar trends than the full 
data-set as expected. The first set is used for the training of the machine 
learning algorithms, while the validation-set is used to estimate the 
performance of each technique.

We also randomly choose one galaxy cluster and its associated 
galaxies (1041 galaxies in the cylinder), 
which are not included neither in the training-set 
nor in the validation-set, but are used to test the final algorithm instead. 
Taking into account that the galaxies in the validation-set are never 
`seen' by the trained machine learning algorithms, we avoid over-fitting the 
data, thus achieving a reliable measurement of the performance of each method 
as explained below.

\subsection{Class-probability of galaxies in the PPSD: a comparison of 
the three ML methods}
Once the ML methods have been trained, we use them to predict, for each galaxy
in the validation-set, the class-probability, $p_i$ ($i=1,\ldots, 5$),
of belonging to a particular class. These predicted probabilities allow us to 
classify galaxies, to measure the performance of each method, and to 
compare their outputs.

\begin{figure*}
\centering
  \includegraphics[width=2.1\columnwidth]{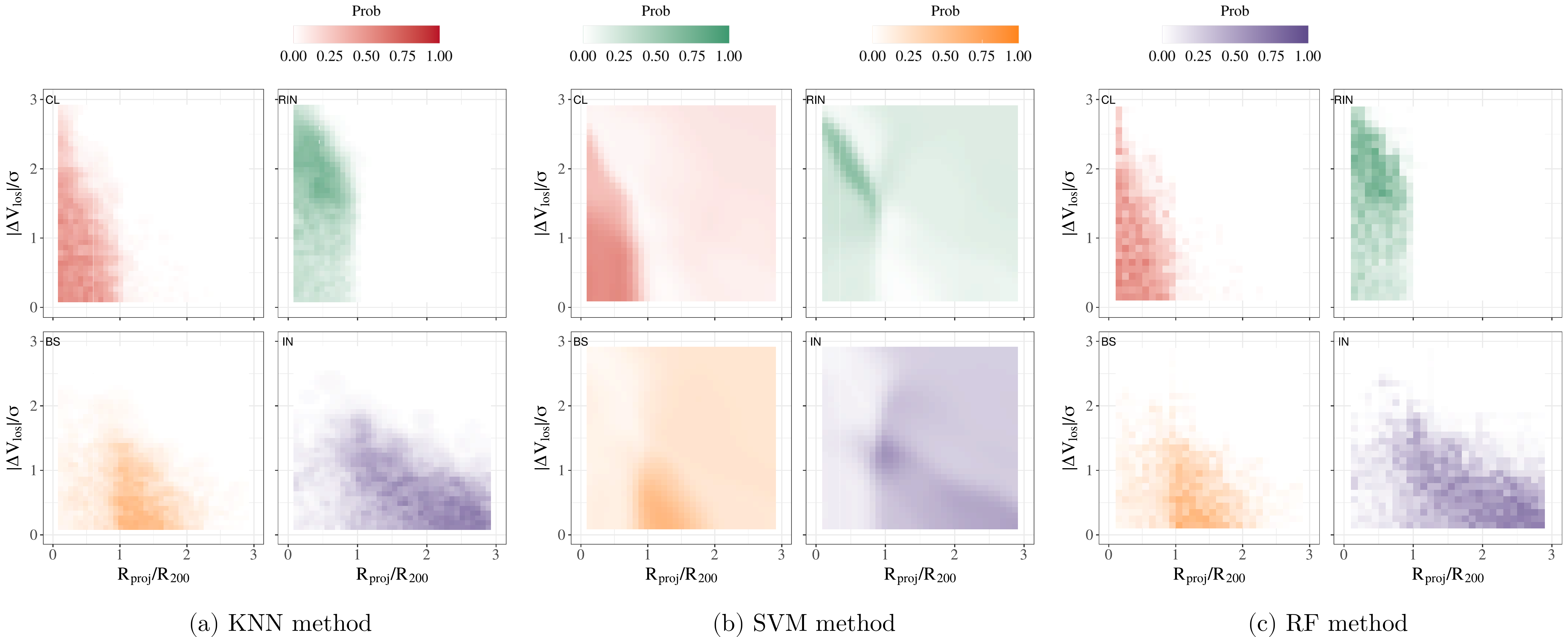}
  \caption{
  Mean class-probability in the PPSD for galaxies in the validation-set.
  Panel (a) shows the results obtained with the KNN method, panel (b) SVM 
  method, and panel (c) RF method. In each case, different sub-panels 
  correspond to galaxies classified as CL (red), RIN (green), BS (orange), and 
  IN (violet).}
  \label{fig:phase_space_probability}
\end{figure*}

In Fig. \ref{fig:phase_space_probability} we show, for each method, 
the mean probability of the four classes of interest (CL, RIN, BS, and IN) as 
a function of the position in the PPSD.  To compute these maps, we perform a
two-dimensional binning of the PPSD and compute the mean value of the 
probability of a particular class for all validation-set galaxies in each bin. 
It can be seen that, as expected, the high-probability regions found by all
methods agree well with the high-density regions of Fig. \ref{fig:phase_space}. 
It is interesting to note that the SVM method assigns a non-zero (however 
very low) probability of being a galaxy of classes (i) to (iv), to regions of 
the phase-space where there are almost exclusively interlopers (class (v)).

\subsubsection{Classification from the class-probability and testing the ML 
methods}
With the estimated class-probabilities, there are 
at least three straightforward schemes for classifying a galaxy:
\begin{description}
    \item 
        (a) The class is given by the highest class-probability.
    \item 
        (b) The class is given by a random pick of the five classes taking into 
        account the estimated class-probabilities. Briefly, given a galaxy with
        class probabilities, $p_i$, where $\sum_{i=1}^{5}p_i=1$, 
        we choose the class using the R-function \texttt{sample} 
        setting the parameter \texttt{prob} $ = p_{i}$.
    \item 
        (c) The class is given by the class-probability that is higher than a 
        certain threshold.
\end{description}

Another utility of the class-probabilities is their use in weighted 
statistics. There might be situations in which it is not desirable to split 
galaxies in the sample under study into classes, but to use all 
galaxies in statistics which involve a weighting scheme. The estimated class-probabilities
can be used as such weights.

In order to compare the performance of the ML methods, we define the 
following statistics: 

\begin{figure}
\centering
  \includegraphics[width=1.1\columnwidth]{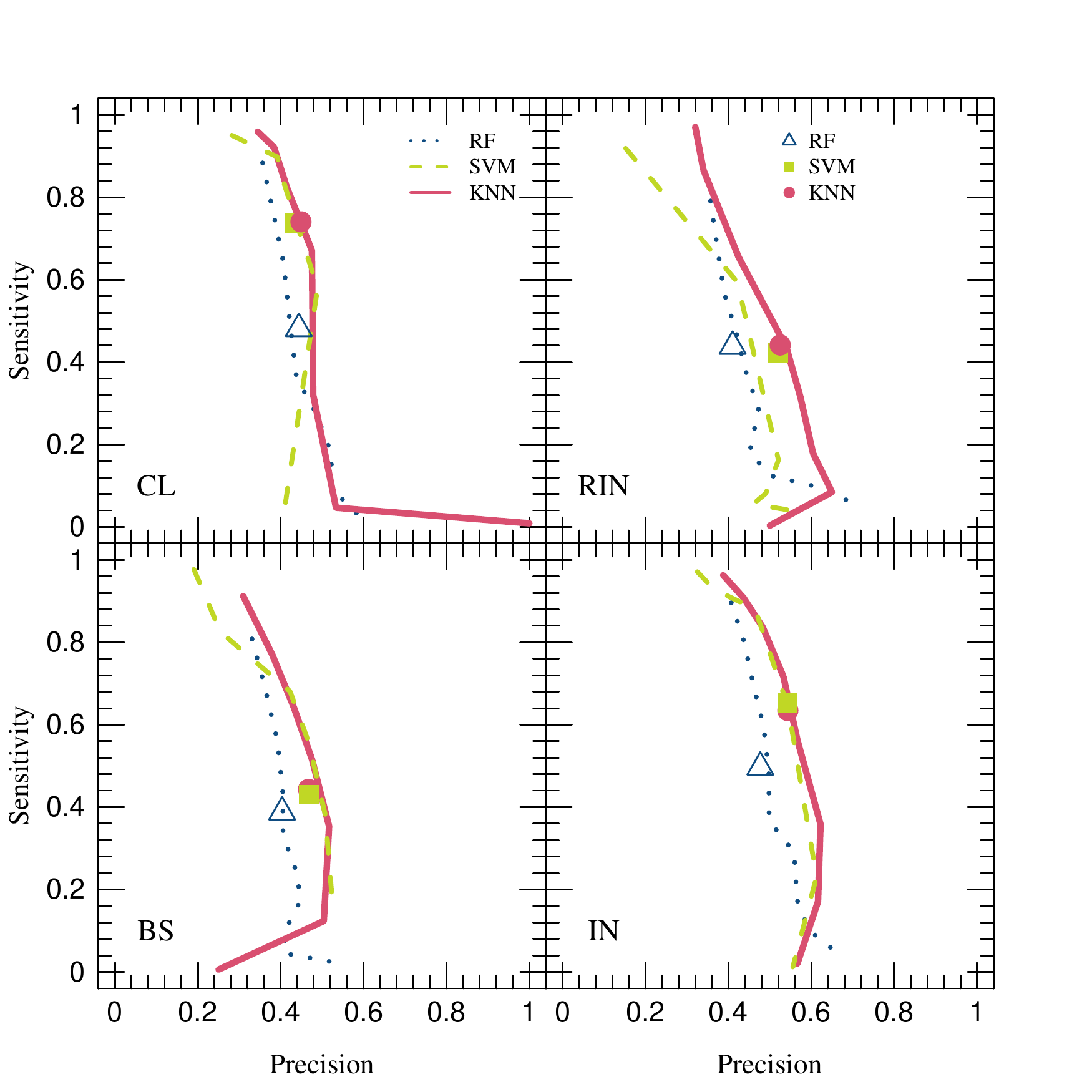}
  \caption{Sensitivity vs. precision of the trained ML algorithms
  represented by different curves. Results are obtained using the scheme 
  (c) that considers a probability threshold as a parameter.
  In the curves shown, the parameter ranges from 0.1 at the upper
  end to 0.9 at the lower end.
  We show as blue triangle, light-green square and
  red filled circle the sensitivity and precision obtained when classifying 
  galaxies taking into account the class with the highest probability. 
  Each panel corresponds to a different class of galaxy in the training-set:
  cluster galaxies (upper left panel), recent infallers (upper right panel) 
  backsplash galaxies, (lower left panel), and infalling galaxies (lower right
  panel).}
  \label{fig:comp}
\end{figure}

\begin{figure*}
\centering
  \subfigure[KNN method]{\includegraphics[width=0.8\columnwidth]{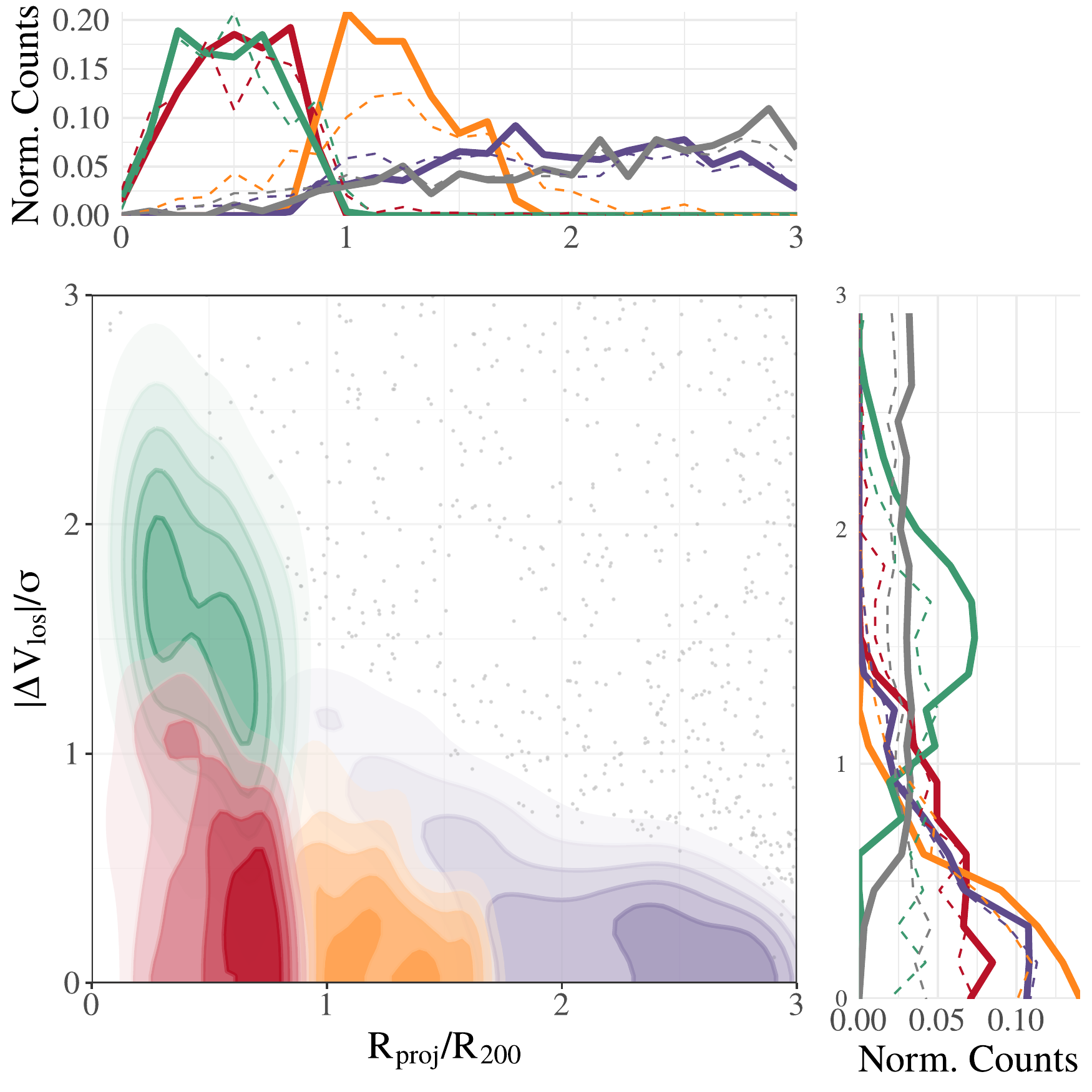}
  \includegraphics[width=0.7\columnwidth]{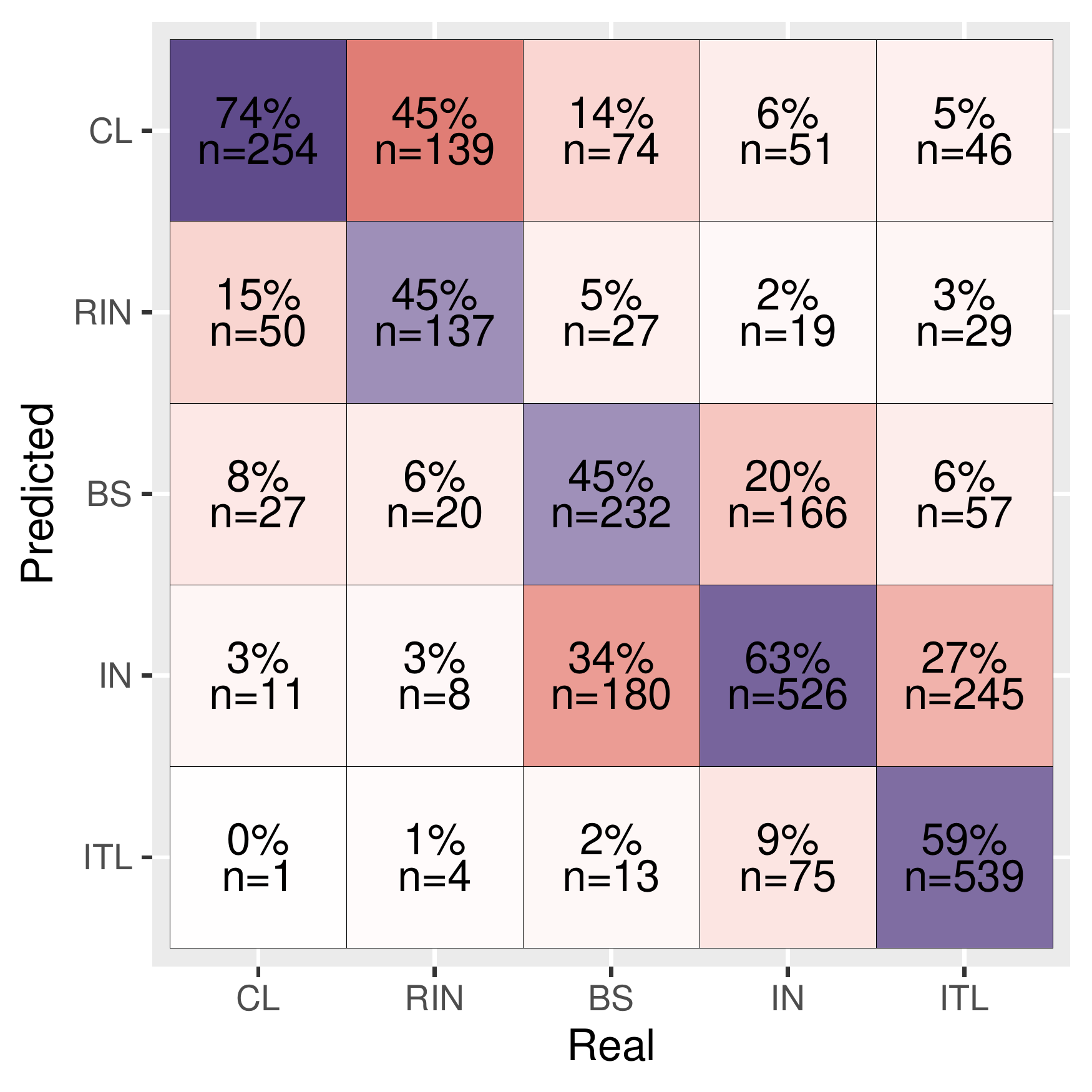}}
  
  \subfigure[SVM method]{\includegraphics[width=0.8\columnwidth]{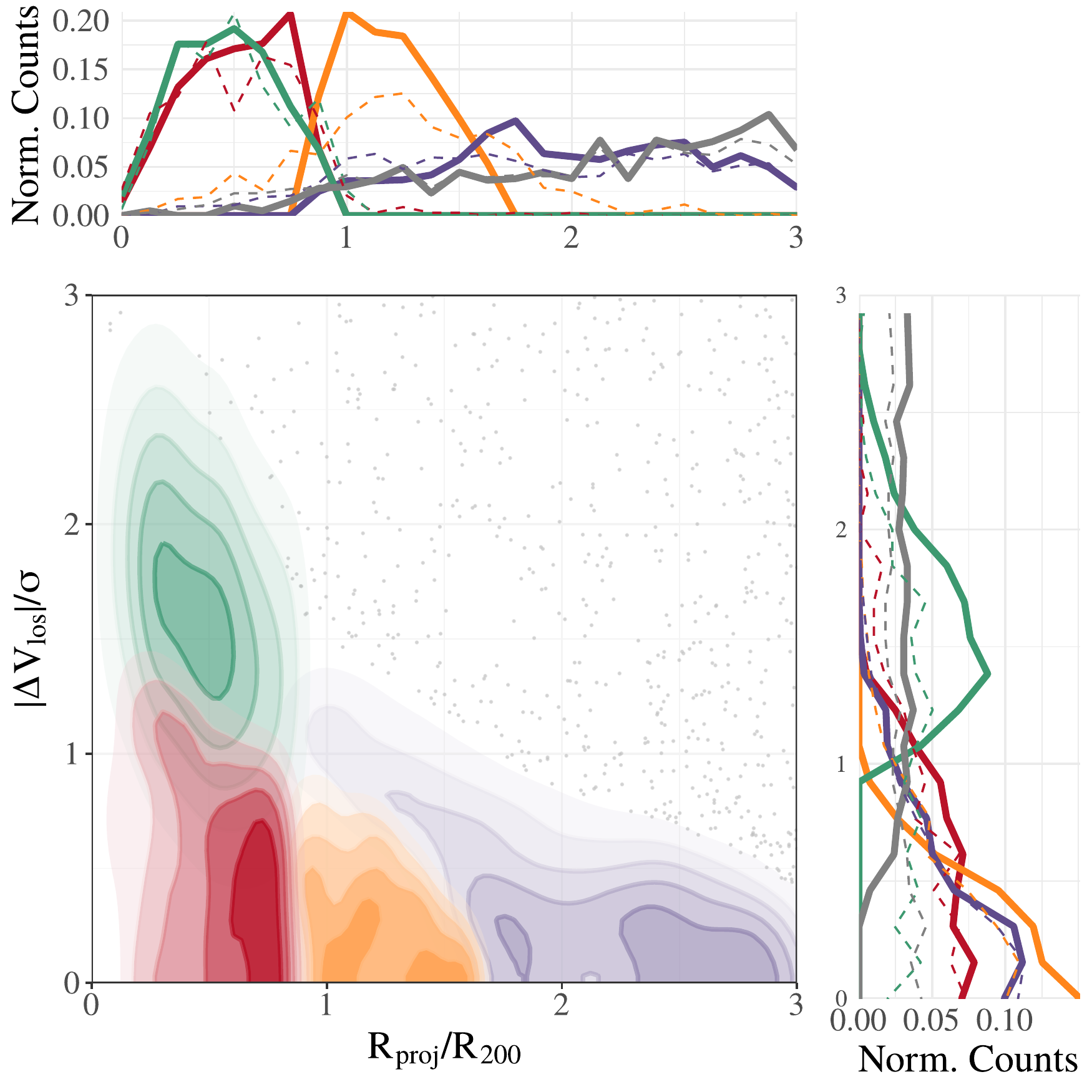}
  \includegraphics[width=0.7\columnwidth]{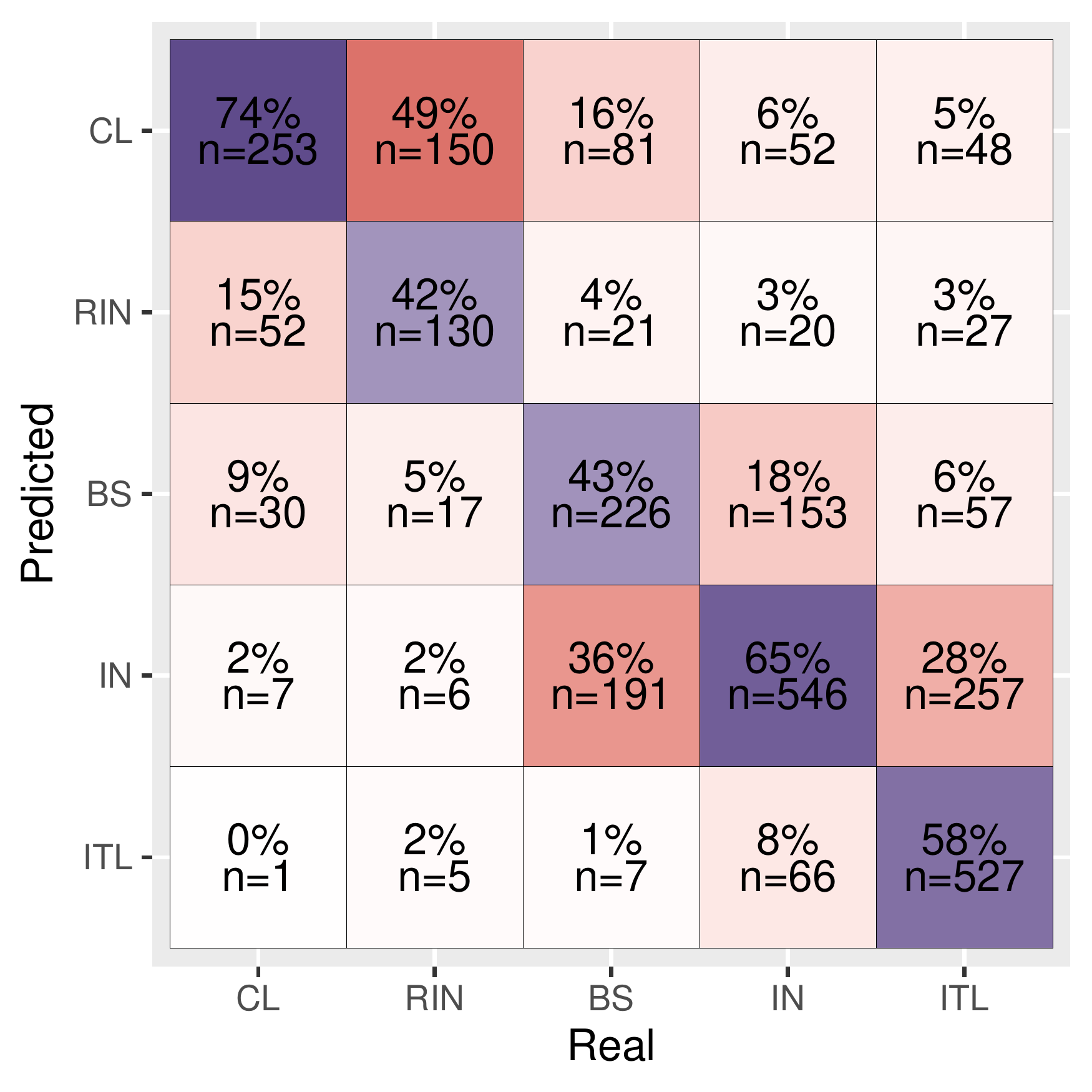}}
  
  \subfigure[RF method]{\includegraphics[width=0.8\columnwidth]{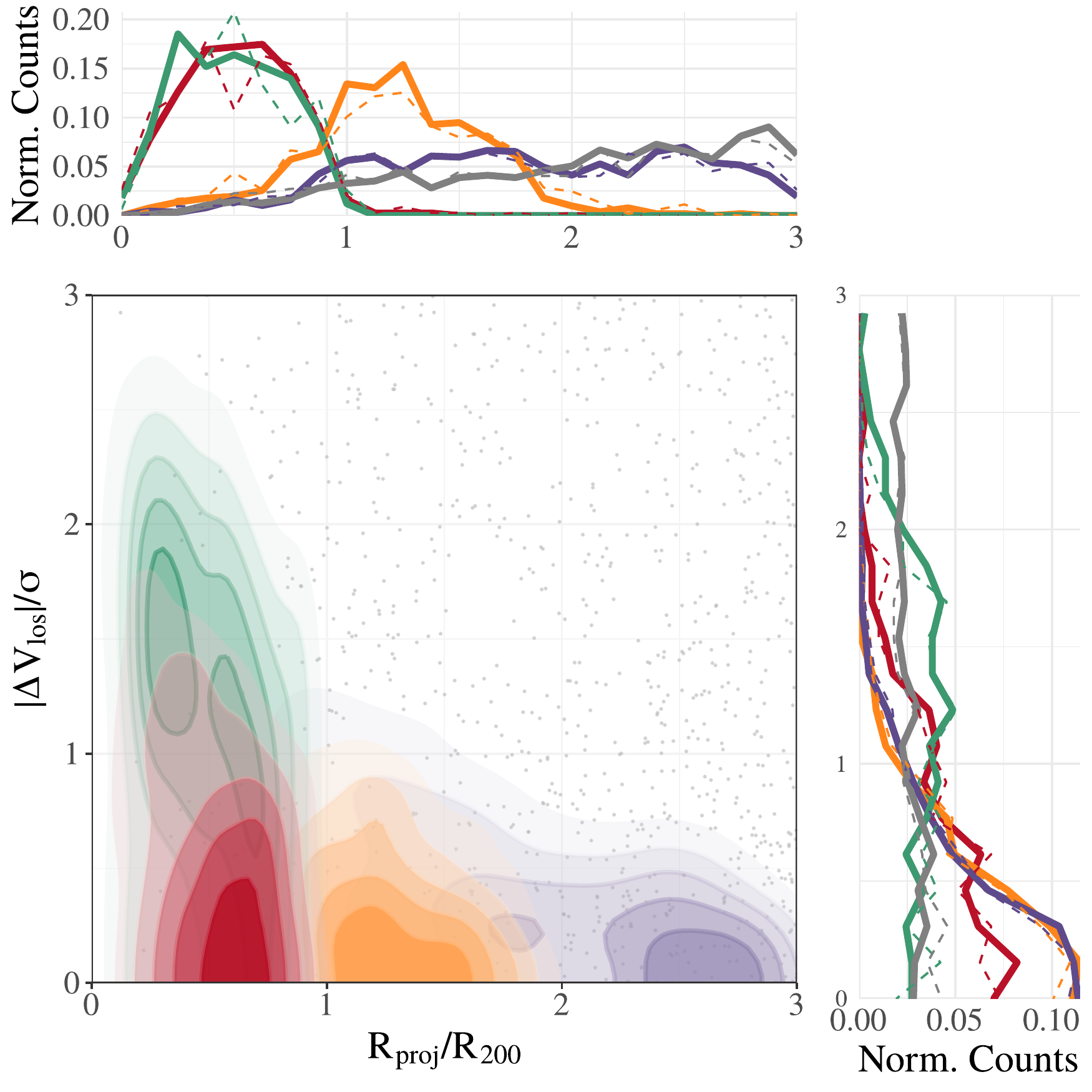}
  \includegraphics[width=0.7\columnwidth]{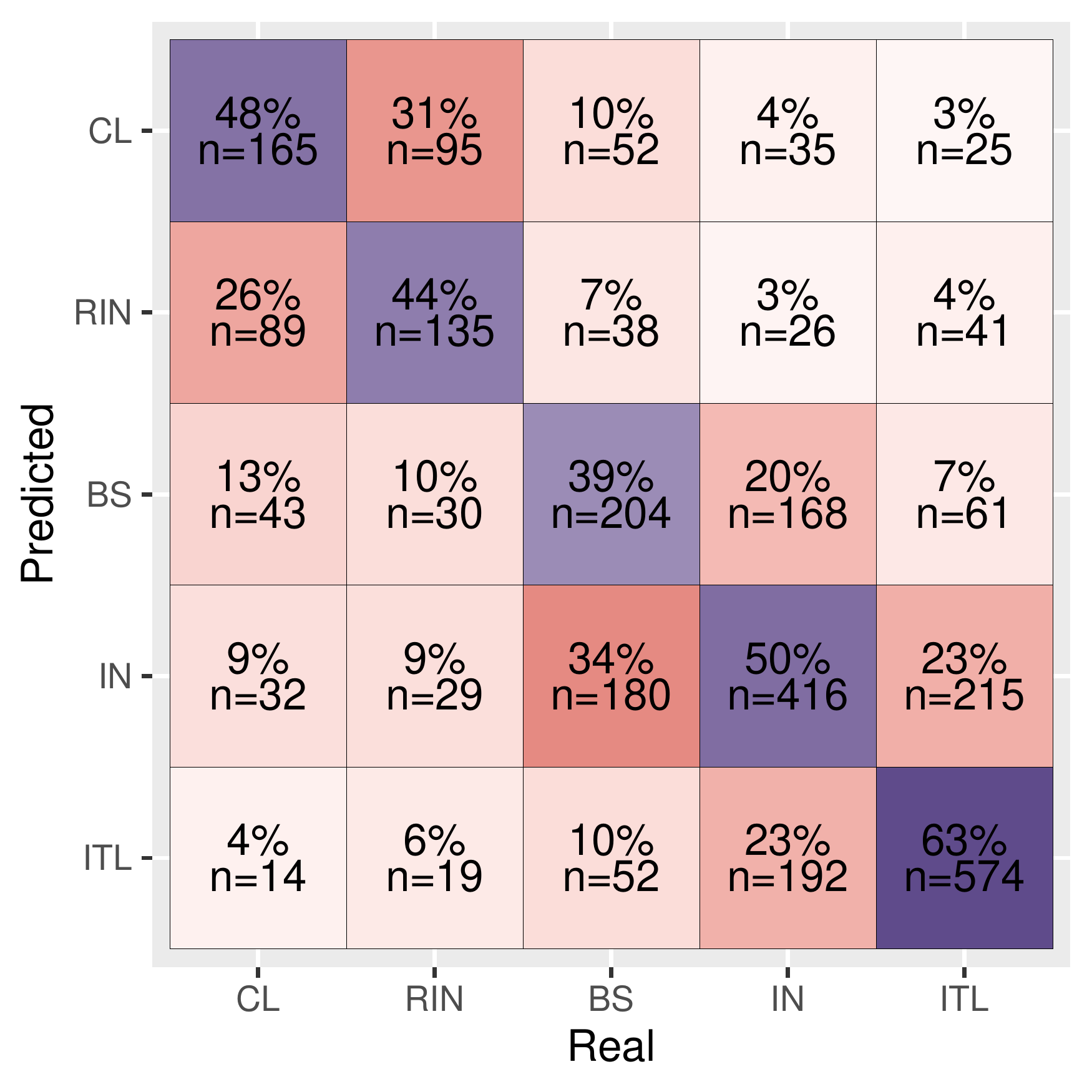}}
  \caption{
  \textit{Left panels}: Phase-space \texttt{ROGER} classification of the galaxies in the
  validation-set determined by the highest class-probability using the three different ML techniques: KNN method (panel a), SVM method (panel b), and RF method (panel c).
 Symbols and colours are as in Fig. \ref{fig:phase_space}. Marginalised 
  normalised distributions are shown at the top and at the right side of each panel
  in continuous lines. We have included as dashed lines the `true' distributions shown in the right panel of Fig. \ref{fig:phase_space}.
  \textit{Right panels}: Confusion matrices associated with the classification of the left panels.}
  \label{fig:phase_space_val_set}
\end{figure*}


\begin{itemize}
    \item Sensitivity: the number of correct predictions of a given class in 
    the resulting sample, divided by the total number of galaxies of that class
    in the validation-set.
    \item Precision: the number of correctly predicted galaxies of a given 
    class in the resulting sample, divided by the total number of galaxies 
    of that class in the resulting sample.
\end{itemize}

Using the scheme (c) above, and taking the 
probability threshold 
as a parameter, we can construct different resulting samples. This procedure, 
in turn, allows us to build a curve in the sensitivity-precision space.
In Fig. \ref{fig:comp}, we show sensitivity vs. precision, for each of the 
trained algorithms, parameterised by the threshold value, that varies from $0.1$ 
(corresponding to the extreme of each curve in the upper left corner
of each panel) to $0.9$ (the other extreme of the curves).
It is important to remark that, given a certain threshold, there can be galaxies 
in the validation-set with all their probabilities lower than this value, and 
so, they will not be classified, and consequently, the sensitivity of the method 
will be reduced. In some extreme cases (see for example the SVM classification
for backsplash galaxies) the are no galaxies with a probability higher than a
threshold of 0.6, making it impossible to compute the sensitivity and precision.
We also show in this figure as a blue triangle, light-green square and
red filled circle the sensitivity vs. precision achieved for each method when 
classifying galaxies taking into account the class with the highest probability
(scheme (a) above).

As expected, the sensitivity decreases as the probability threshold increases.
This happens because increasing the threshold reduces the number of galaxies of
each class in the resulting sample, thus the sensitivity is also reduced.
It is also expected that the precision increases with increasing probability 
threshold. It can be seen that in general this is 
the case, with some few 
exceptions. This can be understood in terms of the overlapping of the different 
classes in the phase-space diagram. Thus there are regions in which, although 
the machine learning methods may compute a high probability for a galaxy of 
being of a certain class, there could also be many galaxies of other classes.
In these regions, an important degree of contamination is expected.

As an example of the ML predictions, we show in the left panels of Fig. 
\ref{fig:phase_space_val_set} the PPSD of the validation-set, where the 
classification of the galaxies is determined by the highest class-probability.
Additionally, we show in solid lines the corresponding
$R_{\rm proj}/R_{200}$ and $|\Delta V_{\rm los}|/\sigma$ marginalised 
distribution for the estimated classification, while in dashed lines we show the 
`true' distributions corresponding to the validation-set (same as right panel of 
Fig. \ref{fig:phase_space}). It can be seen that, although the algorithm is 
capable of recovering similar trends as in Fig. \ref{fig:phase_space}, there 
are regions in which the different classes overlap and so, this echoes in 
contamination on the resulting samples. As discussed above, an important 
feature in Fig. \ref{fig:phase_space} is that some classes overlap more with 
the rest than others. For instance, it is more likely for miss-classified 
cluster galaxies to be classified as recent infallers than as backsplash or 
infallers. This implies that in the resulting predictions the 
miss-classification will not be at random. 

A useful way to visualise the performance of an algorithm is through the 
confusion matrix. Each row of this matrix represents the instances of each 
predicted class, while each column represents the instances of each real class.
On the one hand, from the diagonal of this matrix we can read the sensitivity 
of the method when classifying the different types of galaxies. On the other 
hand, from the off-diagonal terms we can see the miss-classifications.
In the right panels of Fig. \ref{fig:phase_space_val_set}, we present the 
confusion matrices of the classification shown in the left panels, quoting the 
percentages of galaxies of each real class that were classified as 
belonging to each predicted class. For instance, when using the KNN method, 
out of the real CL galaxies, 74 per cent are well classified. The remaining 26
per cent were classified as RIN (15 per cent), as BS (8 per cent), 
as IN (3 per cent) and as INT (0 per cent). The row corresponding to CL galaxies 
shows that 45 per cent of the real RIN galaxies, 14 per cent of the real BS 
galaxies, the 6 per cent of the real IN galaxies, and the 5 per cent of the real 
ITL galaxies were classified as CL. We find that the KNN method achieves a 
sensitivity of 74 per cent, 45 per cent, 45 per cent, 63 per cent, and 59 per cent 
when classifying galaxies as CL, RIN, BS, IN, and ITL, respectively. The SVM
method achieves a 74 per cent, 42 per cent, 43 per cent, 65 per cent, and 58 per 
cent of sensitivity, while the RF method achieves 48 per cent, 44 per cent, 39 per 
cent, 50 per cent, and 63 per cent of sensitivity.
It should be kept in mind that these numbers correspond to the classification 
taking into account the highest class-probability (scheme (a)), and they will 
change if the classification is performed with 
another scheme.
Another feature worth remarking is that, although the RF method recovers 
distributions more similar to the real ones, its performance is poorer
than the SVM and KNN methods in terms of both sensitivity and precision.

Finally, as the \texttt{ROGER} software gives as output the class-probability 
of a galaxy of belonging to each class, we can build a random realisation of a
particular cluster, by randomly classifying  each galaxy as explained in
scheme (b). An example of such procedure is given in Sec. \ref{sec:test}.

Taking into account the results shown in this section, the method of our 
preference is KNN.
It has two main advantages: it is the 
simplest one, and performs similarly or better than the more complex SVM and 
RF methods.  We remark that in the \texttt{ROGER} package 
the three methods are available.

\begin{figure*}
\centering
  \subfigure{\includegraphics[width=1.\columnwidth]{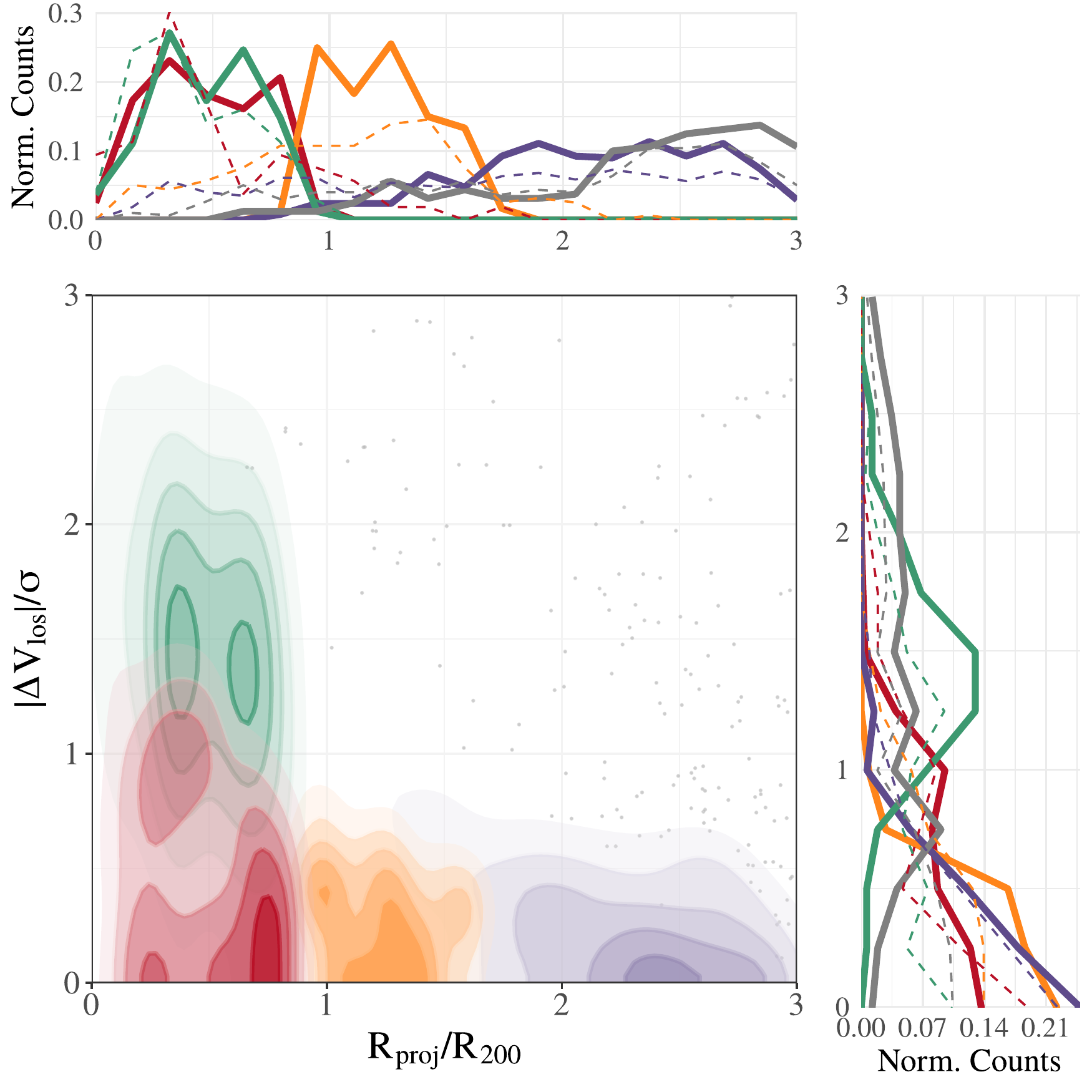}}
  \subfigure{\includegraphics[width=0.9\columnwidth]{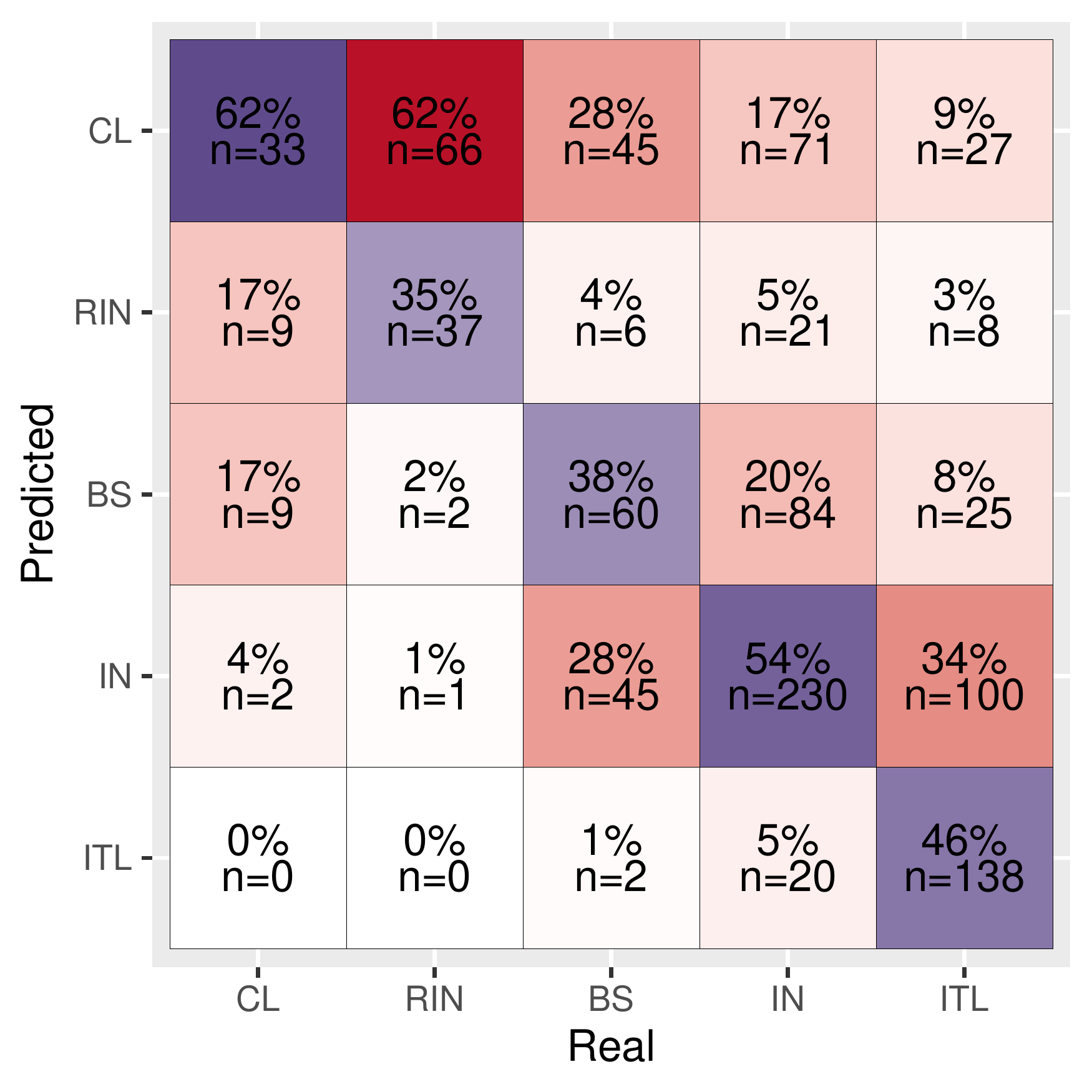}}
  \caption{
  \emph{Left panel}: Phase-space \texttt{ROGER} classification of the galaxies 
  in the test cluster using the KNN method and the highest class-probability for 
  each galaxy. We show in different colour shades the density of galaxies
  classified as CL (red), RIN (green), BS (orange), and IN (violet). ITL 
  galaxies are shown as individual gray points for clarity. 
  In the upper and right sub-panels, we show the marginalised 
  distributions of projected distance and line-of-sight velocity of galaxies, respectively, according to the \texttt{ROGER} classification (continuous lines) 
  and the 'real' classification (dashed lines).
  \emph{Right panel}: confusion matrix associated to the left panel.}
  \label{fig:phase_space_test_set}
\end{figure*}

\begin{figure*}
\centering
  \subfigure{\includegraphics[width=1.\columnwidth]{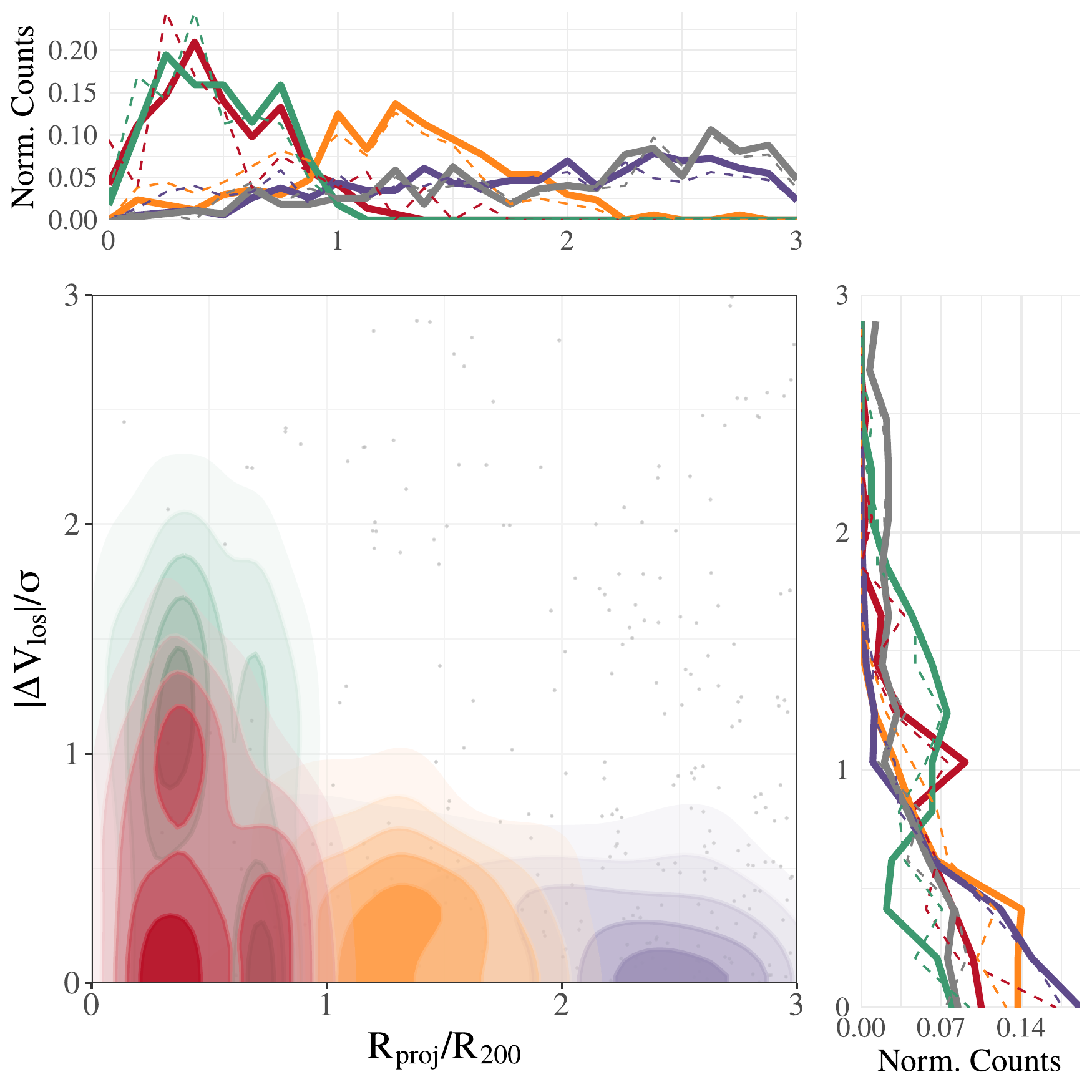}}
  \subfigure{\includegraphics[width=0.9\columnwidth]{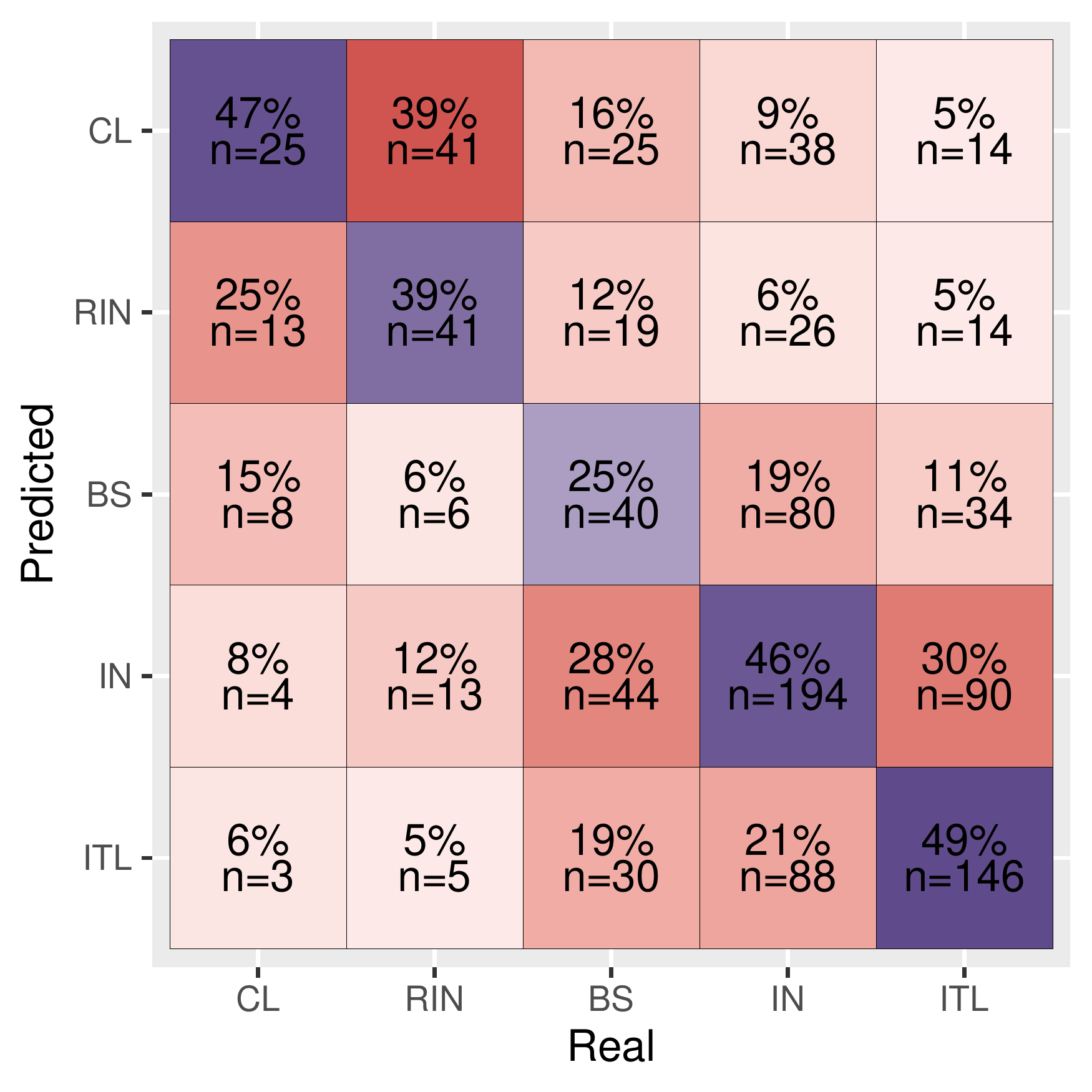}}
  \caption{
  \emph{Left panel}: Phase-space \texttt{ROGER} classification of the galaxies 
  in the test cluster using the KNN method and a random realisation considering 
  the class-probabilities for each galaxy. We show in different colour shades 
  the density of galaxies classified as CL (red), RIN (green), BS (orange), and
  IN (violet). ITL galaxies are shown as individual gray points for clarity.
  In the upper and right sub-panels, we show the marginalised 
  distributions of projected distance and line-of-sight velocity of galaxies, respectively,
  according to the \texttt{ROGER} classification (continuous lines) and the `real' classification
  (dashed lines).
  \emph{Right panel}: confusion matrix associated to the left panel.}
  \label{fig:random_realization}
\end{figure*}

\subsubsection{Testing the methods in an independent cluster} \label{sec:test}
As a further test, and to provide an end-to-end example, we analyse the 
cluster that was previously left aside from both the 
training and validation sets. Several jupyter notebooks with full examples are 
provided on the \href{https://github.com/Martindelosrios/ROGER/}{github 
repository}. We also include an appendix (\ref{sec:apendice}) where we show the
basic options included in the \texttt{ROGER} software.

In the left panel of Fig. \ref{fig:phase_space_test_set}, we show the PPSD of 
the test galaxy cluster, where different colours refer to the types of galaxies
as predicted by the KNN method by taking into account the highest 
class-probability. We show the resulting confusion matrix in the right panel of 
Fig. \ref{fig:phase_space_test_set}. 

Finally, we create a random realisation of the test cluster following the 
scheme (b) as explained in the previous subsection. In the left panel of Fig. 
\ref{fig:random_realization}, we show the PPSD of the random realisation for 
the test cluster and its surroundings. The corresponding marginalised 
distributions are represented by solid lines.
For a better comparison, in dashed lines, we add the marginalised distributions
of galaxy projected distance and line-of-sight velocity
taking into account their real classes. As it can be seen from this figure, 
the distributions in both axis of the random realisation agree well with the 
real distributions of the galaxies, which demonstrate the ability of the method 
to learn the real properties of each galaxy class. In the right panel of Fig. 
\ref{fig:random_realization}, we show the corresponding confusion matrix.
The classification obtained by taking into account the
highest class-probability achieves a better performance in terms of the 
confusion matrix, nevertheless the random realisation reproduces better the
overall distributions.

The results of applying the \texttt{ROGER} algorithm 
to a sample of galaxies completely independent of the samples used for training 
and validation follow the same trends and are consistent with the results 
obtained during the validation process (see Figs. \ref{fig:comp} and 
\ref{fig:phase_space_val_set}). 

\section{Conclusions} \label{sect:conclu}
We have developed \texttt{ROGER}, a machine learning technique-based algorithm 
that classifies
galaxies by relating their position in the projected phase space diagram with 
their three-dimensional orbits. This algorithm was trained using a galaxy 
catalogue generated from the MDPL2 cosmological simulation and the SAG 
semi-analytic model of galaxy formation.
The volume of this simulation is large enough to have a statistically significant
sample of massive isolated clusters and galaxies, which we used to study 
different machine learning methods. For each galaxy, these methods give as 
output the probability of being a cluster galaxy, a recent infaller, a 
backsplash galaxy, an infalling galaxy or an interloper galaxy.
Classifying galaxies into these five classes is useful in studies in which 
it is necessary to know the past trajectory of galaxies, 
in order to understand how different physical mechanisms have acted upon
them. As an example, let us consider a backsplash and a recent 
infaller, they may both be satellites of a cluster,
however, their past 
histories are different. The former has been all the way into the cluster 
and out, while the latter has just dived into the cluster. We should expect 
them to have different physical properties.

Considering a classification scheme that adopts different probability 
thresholds, we were able to build different final samples with different 
contamination and sensitivity. We found that the method with the best 
performance is the K-Nearest Neighbours method, achieving a 74 per cent, 45 
per cent, 45 per cent, 63 per cent and 59 per cent of sensitivity when classifying 
cluster galaxies, recent infaller galaxies, backsplash galaxies, infaller 
galaxies and interloper galaxies, respectively (see Fig. 
\ref{fig:phase_space_val_set}). Although the other methods have similar 
performances and are available in the \texttt{ROGER} software, we 
choose the KNN method as our preferred algorithm for its simplicity.

Finally, with the aim of providing a fully consistent and automatic
algorithm for the classification of galaxies in the PPSD,
we present \texttt{ROGER}, an R-package that is publicly available through the 
github repository: \href{https://github.com/Martindelosrios/ROGER/}
{https://github.com/Martindelosrios/ROGER/}.

\section*{Acknowledgements}
This paper has been partially supported with grants from 
\textit{Consejo Nacional de 
Investigaciones Cient\'ificas y T\'ecnicas} (CONICET, PIP 11220130100365CO) Argentina, the \textit{Agencia Nacional de Promoci\'on Cient\'ifica y Tecnol\'ogica} (ANPCyT), Argentina, and \textit{Secretar\'ia de Ciencia y Tecnolog\'ia, Universidad Nacional de C\'ordoba} (SeCyT), Argentina.
MdlR thanks FAPESP for partial support. HJM thanks H.H. Mart\'inez for useful discussions.
ANR acklowledges funding from ANPCyT (PICT 2016-1975) and SeCyT (PID 33620180101077).
CVM acknowledges financial support from the Max Planck Society through a Partner Group grant. SAC acknowledges funding from {\it Consejo Nacional de Investigaciones Cient\'{\i}ficas y T\'ecnicas} (CONICET, PIP-0387), {\it Agencia Nacional de Promoci\'on Cient\'ifica y Tecnol\'ogica} (ANPCyT, PICT-2018-3743), and {\it Universidad Nacional de La Plata} (G11-150), Argentina.

The \textsc{CosmoSim} database used in this paper is a service by the
Leibniz-Institute for Astrophysics Potsdam (AIP). The authors gratefully
acknowledge the Gauss Centre for Supercomputing e.V. (www.gauss-centre.eu) and
the Partnership for Advanced Supercomputing in Europe (PRACE, www.prace-ri.eu)
for funding the \textsc{MultiDark} simulation project by providing computing
time on the GCS Supercomputer SuperMUC at Leibniz Supercomputing Centre (LRZ,
www.lrz.de).

\section*{Data availability}
The data underlying this article are available in
\href{https://github.com/Martindelosrios/ROGER}{https://github.com/Martindelosrios/ROGER} at
\href{https://zenodo.org/badge/latestdoi/224241400}{https://zenodo.org/badge/latestdoi/224241400}


\bibliographystyle{mnras}
\bibliography{references} 

\appendix
\section{\texttt{ROGER} examples} \label{sec:apendice}
This is a simple example of the basic use of the \texttt{ROGER} software in 
an R-console.

Once the library is installed\footnote{the installation procedure can be found 
in the github repository}, we begin by loading the \texttt{ROGER} library

\begin{verbatim}
library(`ROGER')
\end{verbatim}

Assuming that the data of a galaxy cluster is loaded in a data-frame 
called `cat' with the projected phase-space information of each galaxy,
that have at least two columns named `r' (projected radius in units of 
$R_{200}$), and `v' (line-of-sight velocity relative to the cluster in units 
of the cluster velocity dispersion $\sigma$),
we can just run the following script to estimate the class-probability
of each galaxy using the trained KNN algorithm.

\begin{verbatim}
pred_prob <- get_class(cat, model = knn,
             type = `prob')
\end{verbatim}

This will give as an output a data-frame with five columns that correspond to 
the five class-probabilities. It is worth to remark that the user can use the 
SVM or the RF method just changing the model option with `svm' or `rf', 
respectively. The user can also change the type of prediction from
`prob' to `class' in order to directly predict the
most probable class, or set a probability threshold value
putting `threshold $= x$' to classify galaxies that
have the corresponding class-probability higher that the
selected value.

\bsp	
\label{lastpage}
\end{document}